\documentclass[conference]{IEEEtran}
\usepackage{array}
\usepackage{graphicx}
\usepackage{amsmath}
\usepackage{algorithmic}
\usepackage{algorithm}
\usepackage{epsfig}
\usepackage{url}
\usepackage{epstopdf}
\usepackage{cite}
\usepackage{setspace}
\usepackage{enumitem}
\usepackage{graphicx,lipsum}
\usepackage{caption}
\usepackage[caption=false]{subfig}
\usepackage{tikz}
\usepackage{pdfpages}

\newcolumntype{P}[1]{>{\centering\arraybackslash}p{#1}} 
\newcolumntype{M}[1]{>{\centering\arraybackslash}m{#1}} 

\newcommand\copyrighttext{%
  \footnotesize \copyright~2015 IEEE. Personal use of this material is permitted. Permission from IEEE must be obtained for all other uses, in any current or future media, including reprinting/republishing this material for advertising or promotional purposes, creating new collective works, for resale or redistribution to servers or lists, or reuse of any copyrighted component of this work in other works.}
\newcommand\copyrightnotice{%
\begin{tikzpicture}[remember picture,overlay]
\node[anchor=south,yshift=10pt] at (current page.south) {\fbox{\parbox{\dimexpr\textwidth-\fboxsep-\fboxrule\relax}{\copyrighttext}}};
\end{tikzpicture}%
}
\begin{document}
\bibliographystyle{IEEEtran}
%

\title{Min-O-Mee: A Proximity Based Network Application Leveraging The AllJoyn Framework}

\author{\IEEEauthorblockN{Hatim Lokhandwala, Srikant Manas Kala, and Bheemarjuna Reddy Tamma}
\IEEEauthorblockA{ Indian Institute of Technology Hyderabad, India\\
Email: [cs13m1018, cs12m1012, tbr]@iith.ac.in}}

\maketitle
\copyrightnotice
\begin{abstract}
Close proximity of mobile devices can be utilized to create ad hoc and dynamic networks. These mobile Proximity Based Networks (PBNs) are Opportunistic Networks that enable devices to identify and communicate with each other without relying on any communication infrastructure. In addition, these networks are self organizing, highly dynamic and facilitate effective real-time communication. These characteristics render them very useful in a wide variety of complex scenarios such as vehicular communication, e-health, disaster networks, mobile social networks etc. In this work we employ the AllJoyn framework from Qualcomm which facilitates smooth discovery, attachment and data sharing between devices in close proximity. We develop \textit{Min-O-Mee}, a Minutes-of-Meeting app prototype in the Android platform, utilizing the AllJoyn framework. Min-O-Mee allows one of the participants to create a minutes-of-meeting document which can be shared with and edited by the other participants in the meeting. The app 
harnesses 
the spatial proximity of participants in a meeting and enables seamless data exchange between them. This characteristic allows Min-O-Mee to share not just minutes-of-meeting, but any data that needs to be exchanged among the participants, making it a versatile app. Further, we extend the basic AllJoyn framework to enable multi-hop communication among the devices in the PBN. We devise a novel routing mechanism that is suited to a proximity centric wireless network as it facilitates data routing and delivery over several hops to devices that are at the fringe of the PBN. 
\end{abstract}

\vspace{-1ex}
\section{Introduction}
\vspace{-1ex}
The primary objective of a Proximity Based Networks (PBN) is to share real-time information of a particular transaction that requires collaboration of participants which are in each other's proximity. An intended realization of this idea is the Min-O-Mee app, which will share the minutes-of-meeting (MOM) with all participants of a meeting in real-time. Likewise, other mobile proximity based information sharing apps will also serve as a proof-of-concept. We illustrate the underlying motivation through a realistic example. Let us imagine going out with a group of friends on a weekend trip. All of us have experienced the inconvenience in keeping track of the expenses incurred by the group during the course of the trip. A solution at their disposal is to share a document over the Internet and constantly update it. However, availability of Internet is a constraint that renders this solution less effective as not all members in the group may have Internet access. An innovative idea would be to leverage the fact 
that they are traveling as 
a \textit{group}, close to each other. Thus the \textit{closeness} or spatial proximity of users and their devices, can be leveraged to create an opportunistic network of the devices which will facilitate real-time data exchange among them. Another benefit of employing a proximity based network (PBN) is that latency inherent in data shared via Internet will not be experienced in a PBN. The idea of sharing real-time data through PBNs can be applied to a variety of domains such as vehicular networks, education networks, e-health, gaming, disaster networks, public outreach programs etc., spawning a plethora of useful application scenarios.   

\vspace{-1ex}
\section{Related Research Work}
\vspace{-1ex}
  
PBNs are \textit{opportunistic networks} (ONs), which are a subclass of MANETs. ONs make use of device-to-device communication along with multi-hop relaying. The most promising feature of ONs is that they are \textit{infrastructure-less} networks \emph{i.e.}, they exploit wireless ad hoc networking techniques to establish communication between devices \cite{onet2}. PBNs leverage the twin features of \textit{coverage extension} and \textit{support extension} that opportunistic networking offers. Through coverage extension, a mobile device which is not in the direct range of Radio Access Network (RAN) infrastructure is connected to it through an intermediate node. Likewise, if a device interface is incompatible with the RAN interface, it can be offered access through an intermediate node which is connected to the RAN. The primary challenge in ONs is reliable routing and delivery of data at the destination. Several data routing and delivery mechanisms are proposed in research studies, 
which essentially rely on multi-hop data transmissions. In \cite {onet} authors devise a partial flooding algorithm by randomly selecting the next-hop neighbor to relay the data. Authors in \cite{onet3} suggest strategies for reliable data delivery and congestion control in the network.
\vspace{-1ex}
 \begin{figure}[htb!]
                \centering
                \includegraphics[width=5cm, height=4.8cm]{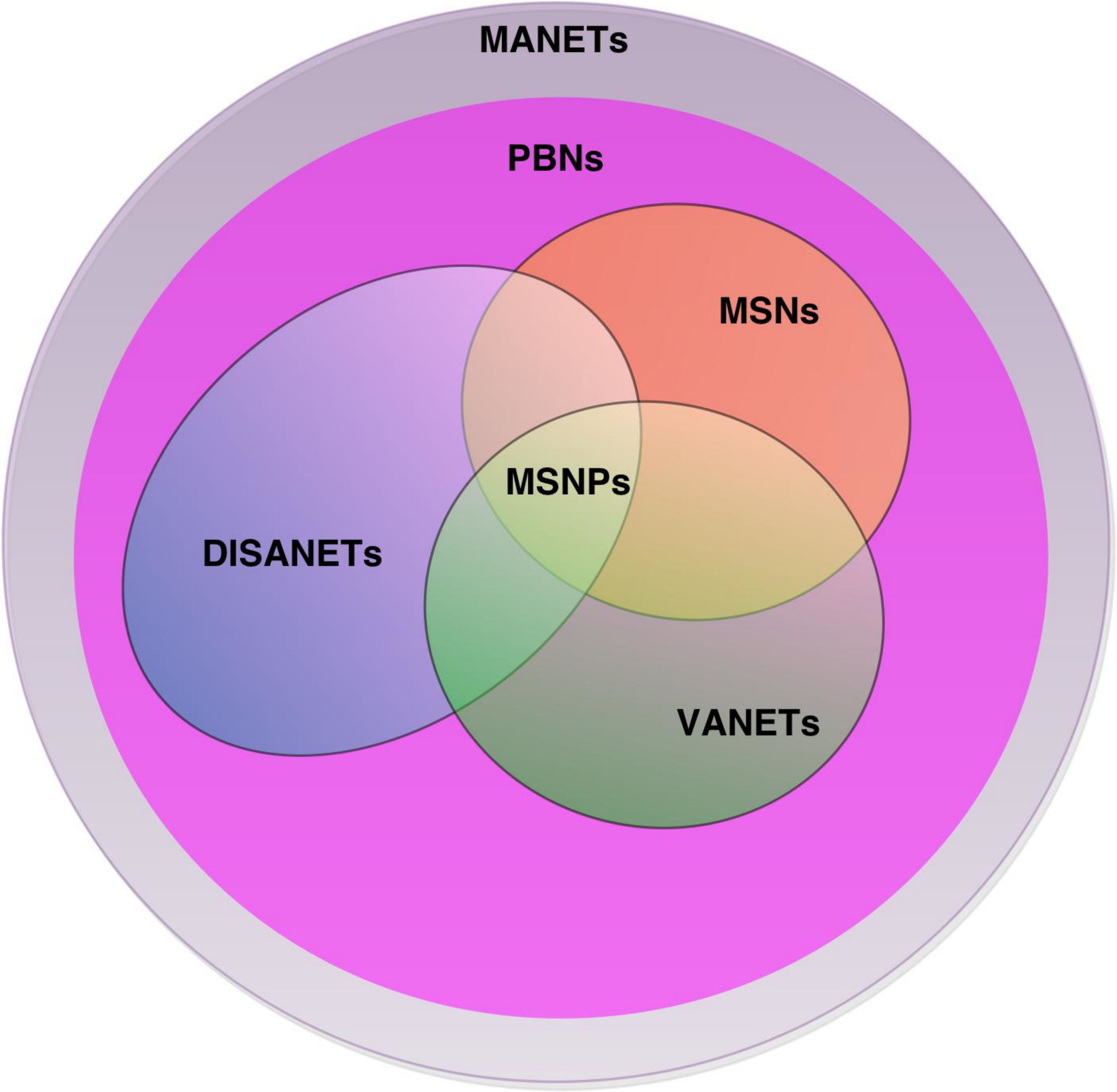}
                \caption{Classification of PBNs}
                \label{clas}
        \end{figure} 
        \vspace{-2ex}
        
PBNs can be further categorized into Disaster Networks (DISANETs), Vehicular Networks (VANETs) and Mobile Social Networks (MSNs) etc., as depicted in Figure \ref{clas}. This classification is based on factors such as, application scenario, network infrastructure, device mobility, data sharing mode \emph{i.e}, real-time or deferred data sharing. PBN architecture forms the basis for mobile social networking in proximity (MSNP). An MSNP application operational over only Wi-Fi Direct is designed by authors in \cite{MSNP2}. A detailed discussion on MSNP architecture and challenges is presented in \cite{MSNP1}. Another mobile PBN application is proposed in \cite{MSNP3}, which focuses on real-time data sharing and human-computer interaction. An MSN based DISANET service for relief and rescue response in urban areas is proposed in \cite{disanet3}. In \cite{vanet}, authors address the challenge of reliable data transmission in highly dynamic and intermittently connected VANETs through an innovative epidemic protocol. 

Despite the variation in problem scenarios and the need of specific infrastructure, the different applications of a PBN share a large number of common challenges \emph{e.g.},  discovery and identification of devices, uninterrupted connectivity, device mobility, reliable data transmission etc. Thus, it is necessary to understand the characteristics of PBNs and then harness these features to tackle the prevalent communication issues.

The rest of the paper is organized as follows. Section III describes about the characteristics of PBNs. Section IV gives an overview about the AllJoyn framework which facilitates developing proximity based applications. In Section V we present a MSNP based application: Min-O-Mee describing its real life usage, functional details and its working model. Section VI covers about the extenstions to the basic framework to realize multi-hop data transmissions and its implementation in the Min-O-Mee application so developed. Section VII covers application deployment and testing followed by conclusion in Section VIII.

\vspace{-1ex}
\section{Characteristics Of Proximity Based Networks}
\vspace{-0.8ex}
Below we elucidate some special features of PBNs:
\vspace{-0.5ex}
\begin{itemize}
 \item \textbf{Spatial Proximity Of Devices:} Closeness of devices is the fundamental characteristic of PBNs which enables them to establish device-to-device communication. A PBN is created when several devices in close proximity are willing to participate in wireless data communication.
 \item \textbf{Multi-hop Data Transmission:} Every device in a PBN forwards the data of which it is not the intended recipient onto the next hop. This ability of the network to relay data over multiple-hops allows it to offer extended coverage, reliable data delivery and resilience to failures.
 \item \textbf{Real-time Data Sharing:} Information can be instantaneously shared with all the nodes in a PBN. Due to close proximity and multi-hop relaying of data, the latency in data sharing is minimal.
  \end{itemize}

 MSNPs are a subclass of PBNs which are proximity based social networks. However in an MSNP, establishment of connections based on proximity alone will not suffice because a \textit{social link} between any two devices requires \textit{validation} or \textit{trust}. Thus, while a PBN offers a purely anonymous ON service, an MSNP builds a validation layer on top of it. Secondly, the lifespan of a PBN network may be limited to the period in which the PBN physically exists and engages in active communication. An MSNP may have an extended lifespan stretching over multiple sessions of network activity	. Thus, a device which is a recurring member of the MSNP application sessions and has been validated once by other devices, should not have to undergo trust validation exercise upon rediscovery and reattachment to the MSNP.

\begin{figure}[htb!]
                \centering
                \includegraphics[width=8cm, height=5cm]{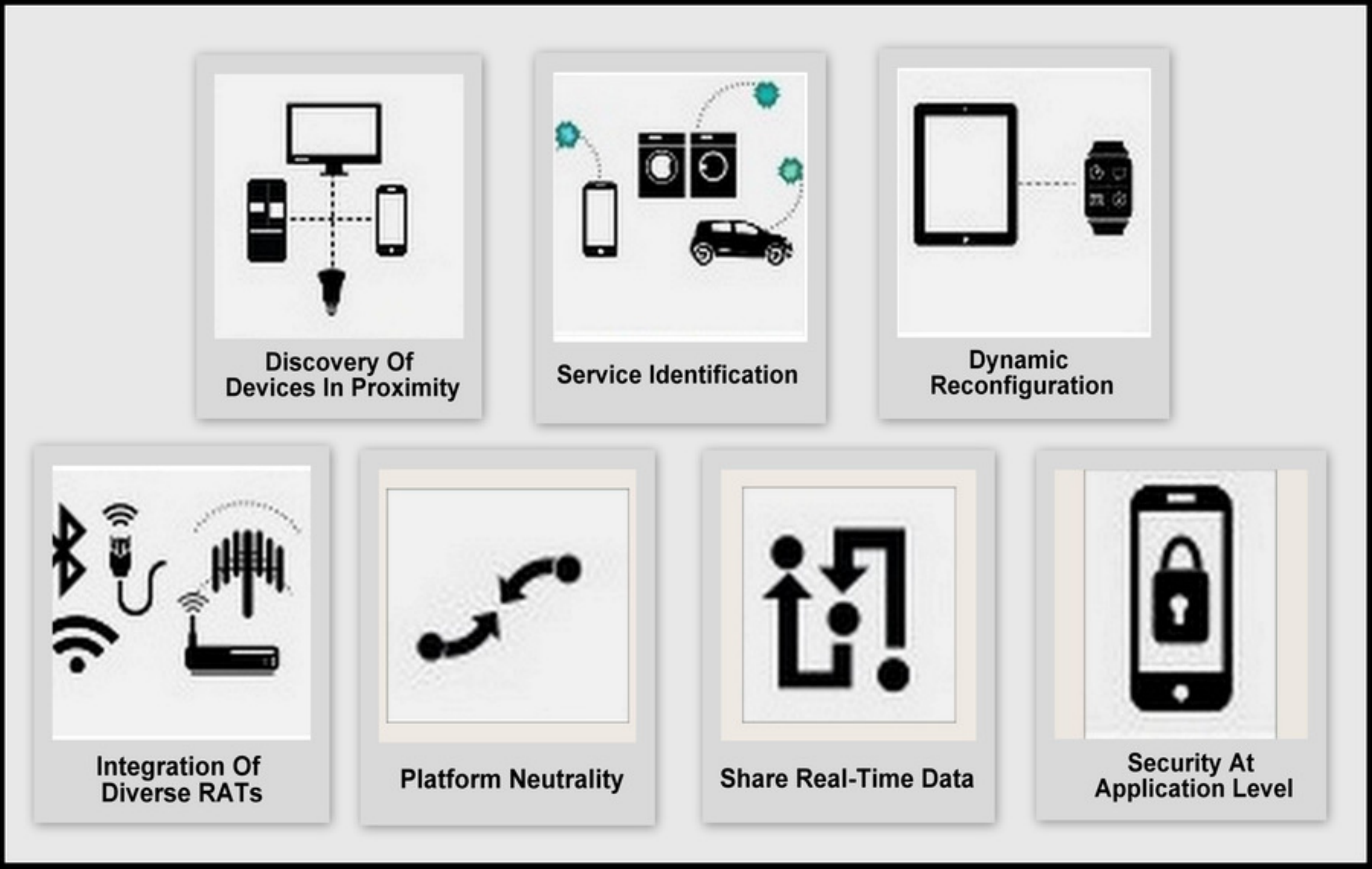}
                \caption{Features Of The AllJoyn Framework}
                \label{features}
        \end{figure} 
 \vspace{-1ex}
\section{The AllJoyn Framework}
\vspace{-1ex}
AllJoyn is an open source framework \cite{asa} which is platform-neutral and facilitates proximity based networking across heterogeneous devices using different RATs to communicate. It simplifies automatic discovery of mobile devices in close proximity and establishes communication sessions between them. The framework offers several advantages that were lacking in the domain of proximity network development. The features of the AllJoyn framework are illustrated in Figure~\ref{features} and the benefits are presented below.
\begin{itemize}
 \item The AllJoyn framework is an open source project and hence is useful for research community as well as for commercial ventures.

 \item The AllJoyn framework offers a device discovery and service advertisement mechanism as an abstraction which operates across heterogeneous technologies.
  \item It allows dynamic configuration of the network.
 \item AllJoyn implementations can execute on multiple operating systems, making it platform independent.
 \item Developers can pick programming language of their choice (C, C++, Java etc.) for developing applications.
 \item It provides independence from lower layer network stacks through smart interfaces which a developer can easily work with.

 \item It provides greater security by allowing access at the granularity of application-to-application communication.
\end{itemize}

Thus, the AllJoyn framework provides an easy-to-implement object model and is a comprehensive open-source tool for development of network applications which are proximity centric.
\vspace{-1ex}
\section{Min-O-Mee : A Minutes Of Meeting App}
\vspace{-1ex}
We now present Min-O-Mee, a proximity based application designed to record the minutes-of-meeting (MoM/MOM). The objective of the application (App) is to share the MOM with the participants of a meeting in real time. The AllJoyn framework is ideal to develop Min-O-Mee as all participants of a meeting will be in close proximity. Min-O-Mee is built on top of the AllJoyn framework abstractions, which provide the services such as session management, participant device discovery and attachment, and data transfer. In the following subsections, we describe the basic App design, detailed functionalities, smart features, extension to the basic framework and planned enhancements.

\vspace{-1ex}        
\subsection{\textbf{Basic Design}}
\vspace{-1ex}
\begin{itemize}
 \item There will be two types of participants in a meeting. 
 \begin{itemize}
  \item \textit{Scribe} : A single participant who will create the MoM.
  \item \textit{Member(s)} : The remaining participants who would view and receive the MoM.
 \end{itemize}
 \item Scribe initiates the Min-O-Mee session and advertises a unique \textit{session name}.
 \item The Scribe prepares the MOM which could be shared with the Members in two ways.
 \begin{itemize}
  \item \textit{Real-Time Sharing } : A \textit{text-pad} on each Member device is instantaneously updated as the Scribe prepares the MoM. Member devices can view the text-pad in the read-only mode.
  \item \textit{Deferred Sharing} : The Scribe sends a copy of the MoM as a text/PDF file to all the Member devices immediately after the meeting is over. 
 \end{itemize}
\end{itemize}
\vspace{-2ex}
\subsection{\textbf{Functional Details}}
\vspace{-1ex}
We elaborate upon the functional details, by dividing the App functions into five operational categories \emph{viz.}, Peer Advertisement And Discovery, Services and Application Architecture, App Dashboard, Text Editor and Data Sharing. 

\subsubsection{Peer Advertisement And Discovery}
All participants advertise a unique device ID which comprises of a \textit{social} component and an \textit{identification} component. The device \textit{name} is the social component which the user interface (UI) prompts the Scribe/Member to enter when he/she starts the application. It is the device advertisement mechanism of Min-O-Mee and the device name should ideally be the Member's name as other Members would easily recognize it. But human names are not unique and are unsuitable to be device IDs. We remedy this problem by concatenating the device name with the subscriber identification module (SIM) number of that particular device. The purpose of this identification component is to transform the device name into a unique device ID. While the Members only see the social device names, the App running on every device receives the unique device IDs of other Members.  
        \begin{figure}[htb!]
                \centering 
                \includegraphics[width=5.3cm, height=6.2cm]{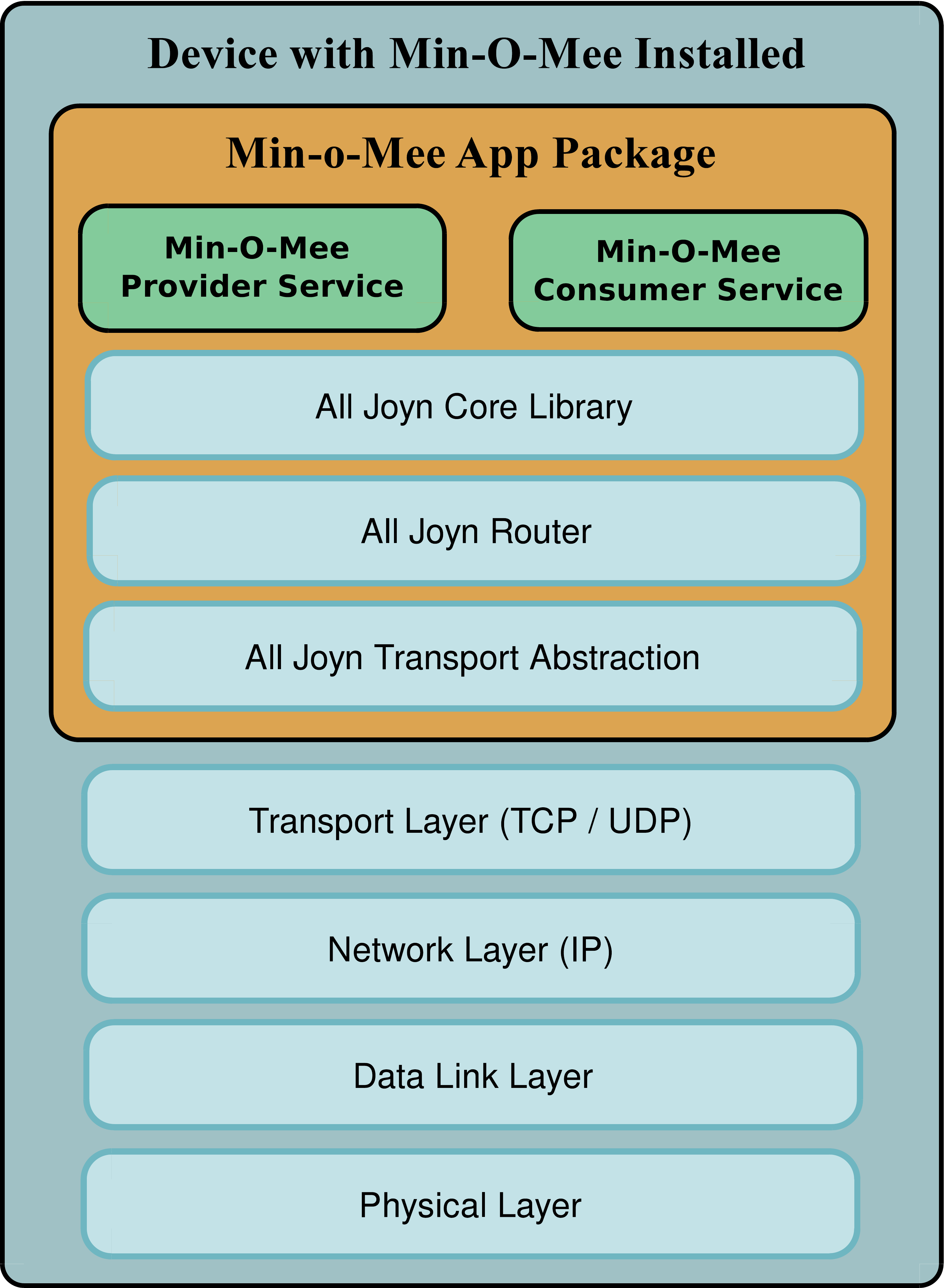}
                \caption{AllJoyn Application Architecture}
                \label{Arch}
        \end{figure}
        
\subsubsection{	Services and Application Architecture}
In the context of the AllJoyn terminology each AllJoyn application comprises of multiple services. These services could be consumer services or provider services. Consumer services consume services provided by the other applications whereas provider services provides services to the other applications in the proximity. For E.g. consider a motion sensor that determines whether there was movement of people or objects within its range. The motion sensor could be hosting a provider service that intimates the near by devices whenever a motion was detected. A light bulb in the vicinity having an AllJoyn application with some consumer service would listen to the data provided by the producer service (in this case the motion sensor). Based on the received data the light bulb can turn on/off. Thus here motion sensor acts as a service provider and the light bulb acts as service consumer. An AllJoyn application may have consumer service(s) or producer service(s) or a combination of both. 

Min-O-Mee application that we develop has both a producer and a consumer service embedded within it. Figure \ref{Arch} illustrates where the application resides in the protocol stack and modules within the application package. Producer service within the application lets a device advertises about its capabilities through advertisement mechanism of hosting a Min-O-Mee session. Consumer service on the other hand lets a device join a Min-O-Mee session offered by an application on some other device. Advertisement mechanisms lets each device advertise about its producer service with the following details : the path (\textit{OBJECT\_PATH}) on which the service is accessible, the \textit{CONTACT\_PORT}, expected input parameters (if any) to access the service, type of return arguments and metadata related to the application and the device capabilities. Discovery mechanism lets a device discover about the producer services in the vicinity capable of hosting a Min-O-Mee session. 

AllJoyn router in Figure~\ref{Arch} is the module that performs the functions of service advertisement and discovery of remote services. AllJoyn library serves as a medium for the application containing producer and consumer services to interact with the AllJoyn router. Below the router resides the AllJoyn transport module which is an abstraction that hides the details of the underlying transport details, Hardware and the OS details from the app users. In conclusion the entire Min-O-Mee application package resides at the application layer of the standard International Standards Organization Open Systems Interconnection (ISO/OSI) model.

At a time a device could either act as a scribe and provide the services to other or could act as a participant and consume the services provided by some other scribe. For the sake of simplicity and generality, we assume that members in a meeting or a discussion could decide about who amongst them would act as a scribe and the participants through mutual agreement. 
 Accordingly our application provides each user with the list of names of the devices in the vicinity including its own device. Following two cases arise in this context.
 \begin{itemize}
  \item If the user chooses its own name from the list of users/peers in the vicinity then this is considered as in the user wants to act as a scribe. In this case it provides services to the other users, joins the session hosted by itself and let other users join the session.
  \item If the user chooses some other user name from the peer list then it is considered that the user wants to act as a participant in the meeting/discussion. Here the user joins the session hosted by the selected peer.
 \end{itemize}

There can be multiple scribe(s) at a particular point of time but a participant could join session hosted by only single scribe at a time. If the participant opts to join the session hosted by some other scribe, he/she would be automatically disconnected from the session it was previously the part of. 
\subsubsection{App Dashboard} The App Dashboard is divided into two sections, \emph{viz.}, 'My MoMs' and 'Shared MoMs'. 'My MoMs' is the list of MoMs which are created by that particular member \emph{i.e.}, she was the Scribe in the meetings whose MoMs are listed. 'Shared MoMs' is the list of MoMs that were shared with the particular member when some other member was the Scribe.

\begin{figure}
  \begin{tabular}{cc}
   \subfloat['My MoMs' List ]{\includegraphics[width=4.25cm, height=6cm]{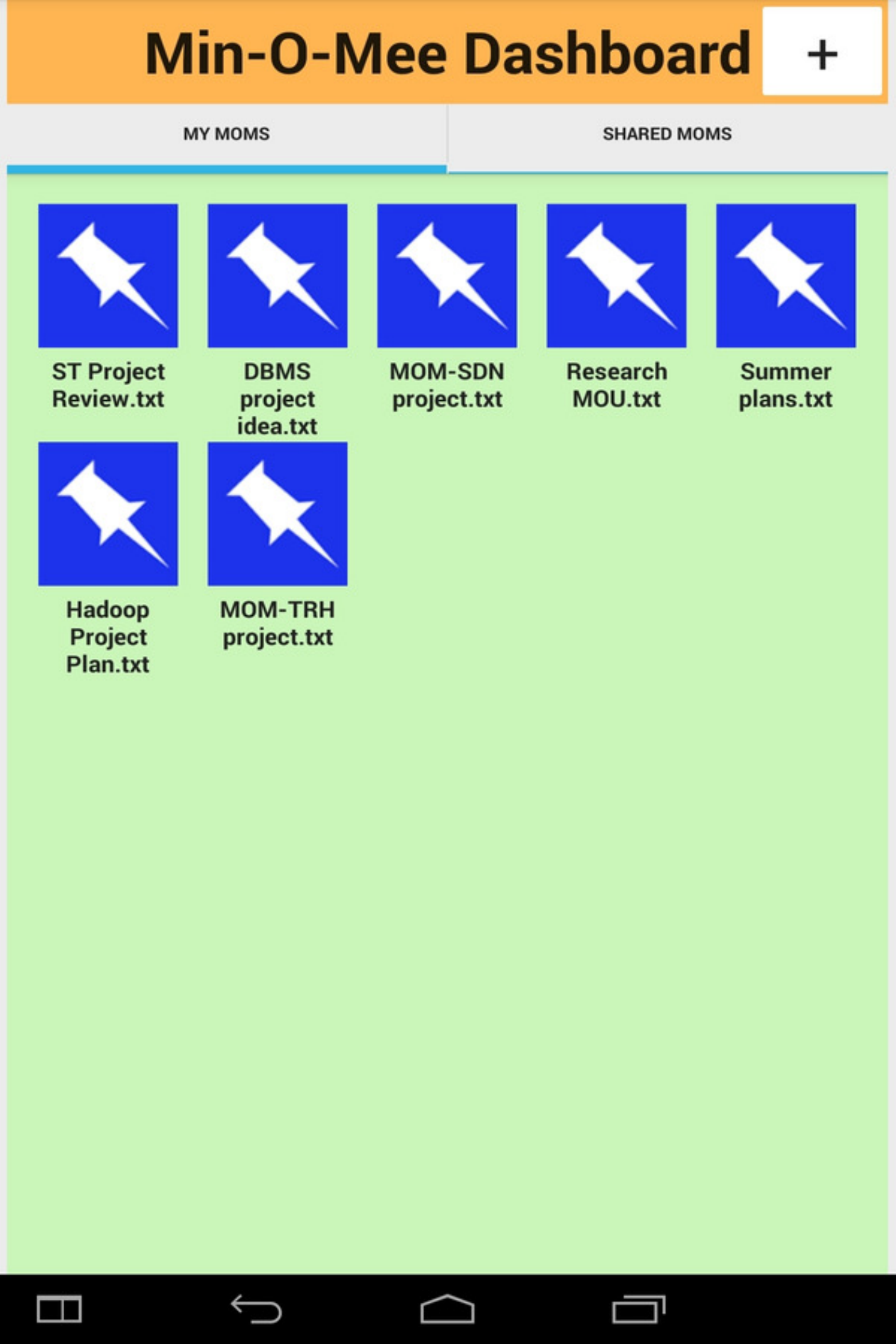}}\quad%
   \subfloat['Shared MoMs' List] {\includegraphics[width=4.25cm, height=6cm]{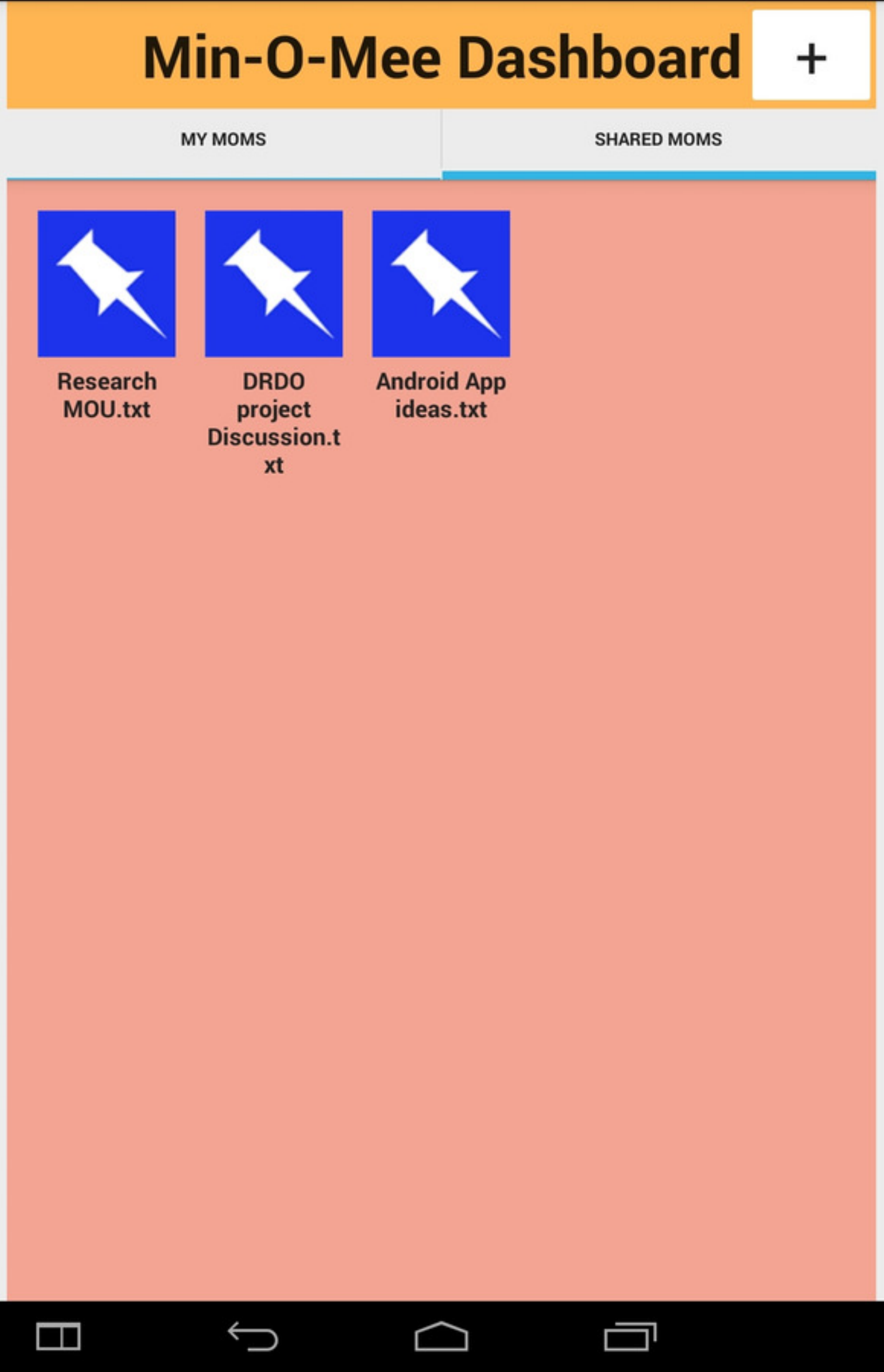}}%
   \end{tabular}
    \caption{The Min-O-Mee Dashboard} 
     \label{ss2}
\end{figure}

\begin{figure}
  \begin{tabular}{cc}
   \subfloat['My MoMs' File Options ]{\includegraphics[width=4.25cm, height=3.6cm]{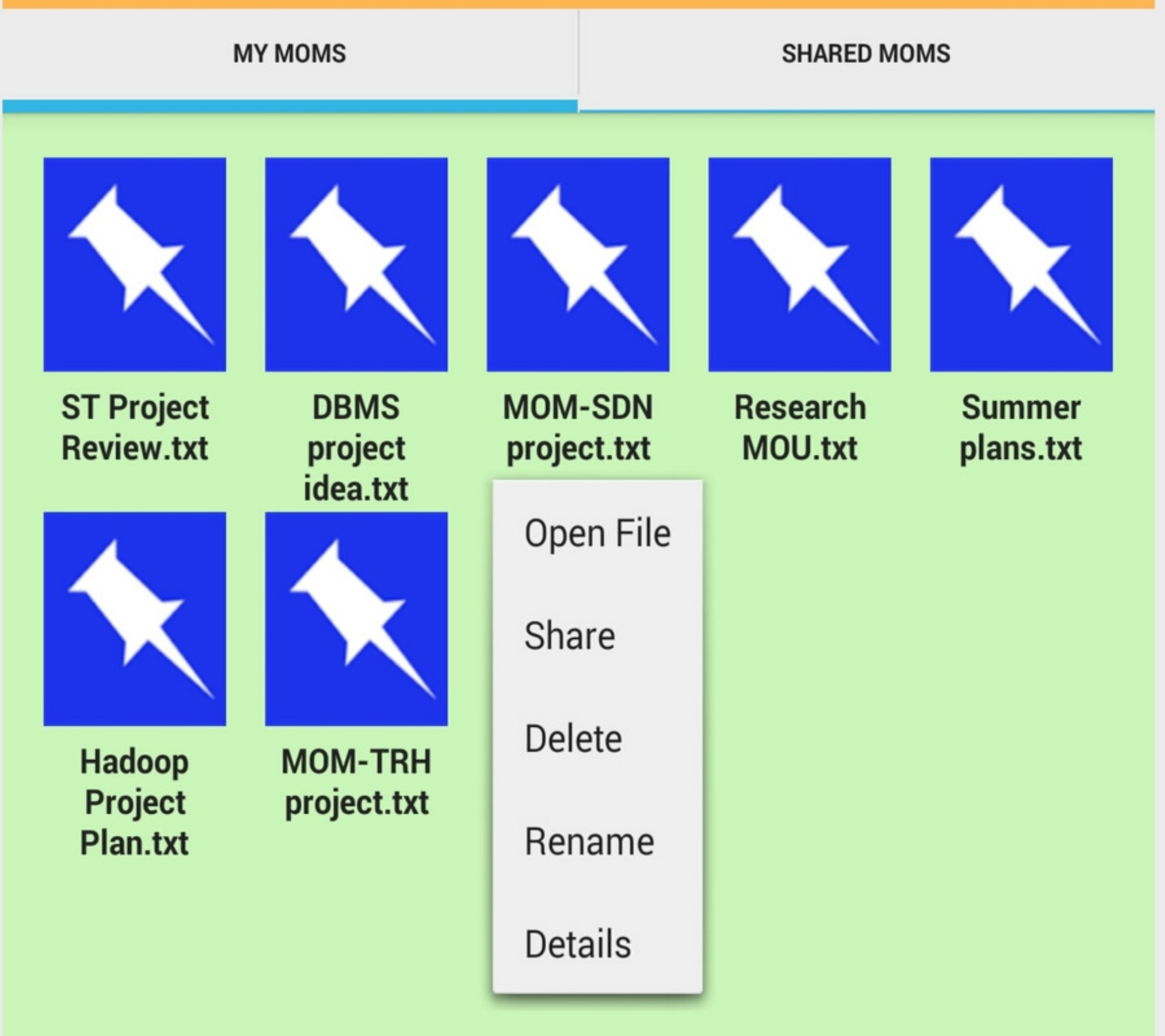}}\quad%
   \subfloat['Shared MoMs' File Options] {\includegraphics[width=4.25cm, height=3.6cm]{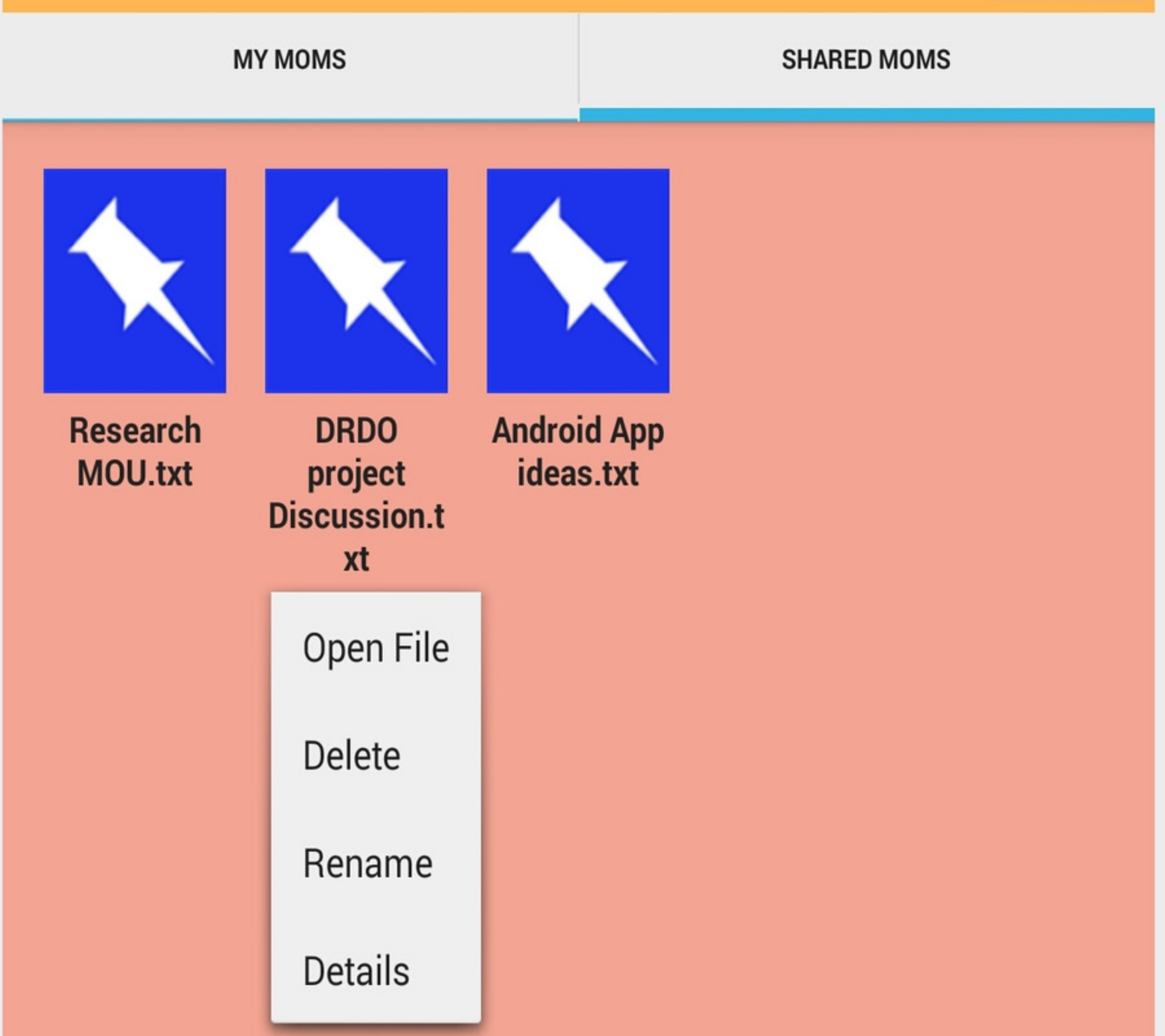}}%
   \end{tabular}
    \caption{Difference In File Options In The Two Lists} 
     \label{ss3}
\end{figure}

A screenshot of the Min-O-Mee Dashboards of two different devices is exhibited in Figure~\ref{ss2}. In Figure~\ref{ss2}~(a), the 'My MoMs' tab list the MoMs from the 'My MoMs' folder on a certain device (Lenovo Yoga Tablet). Similarly in Figure~\ref{ss2}~(b), the 'Shared MoMs' tab list the MoMs from the 'Shared MoMs' folder in an another device (Samsung Galaxy DUOS I9082). The choice of the colors also highlights the fact that the files in 'My MoMs' list can be edited by the user but she can not modify the content of the files in the 'Shared MoMs' list. However, files in both the lists can be read, renamed and deleted by the respective users. The various file operations that are permissible in the two lists are illustrated in Figure \ref{ss3}. 

\subsubsection{Text Editor}
\begin{figure}
  \begin{tabular}{cc}
   \subfloat[Viewing/Editing 'My MoMs' File]{\includegraphics[width=4.25cm, height=6.5cm]{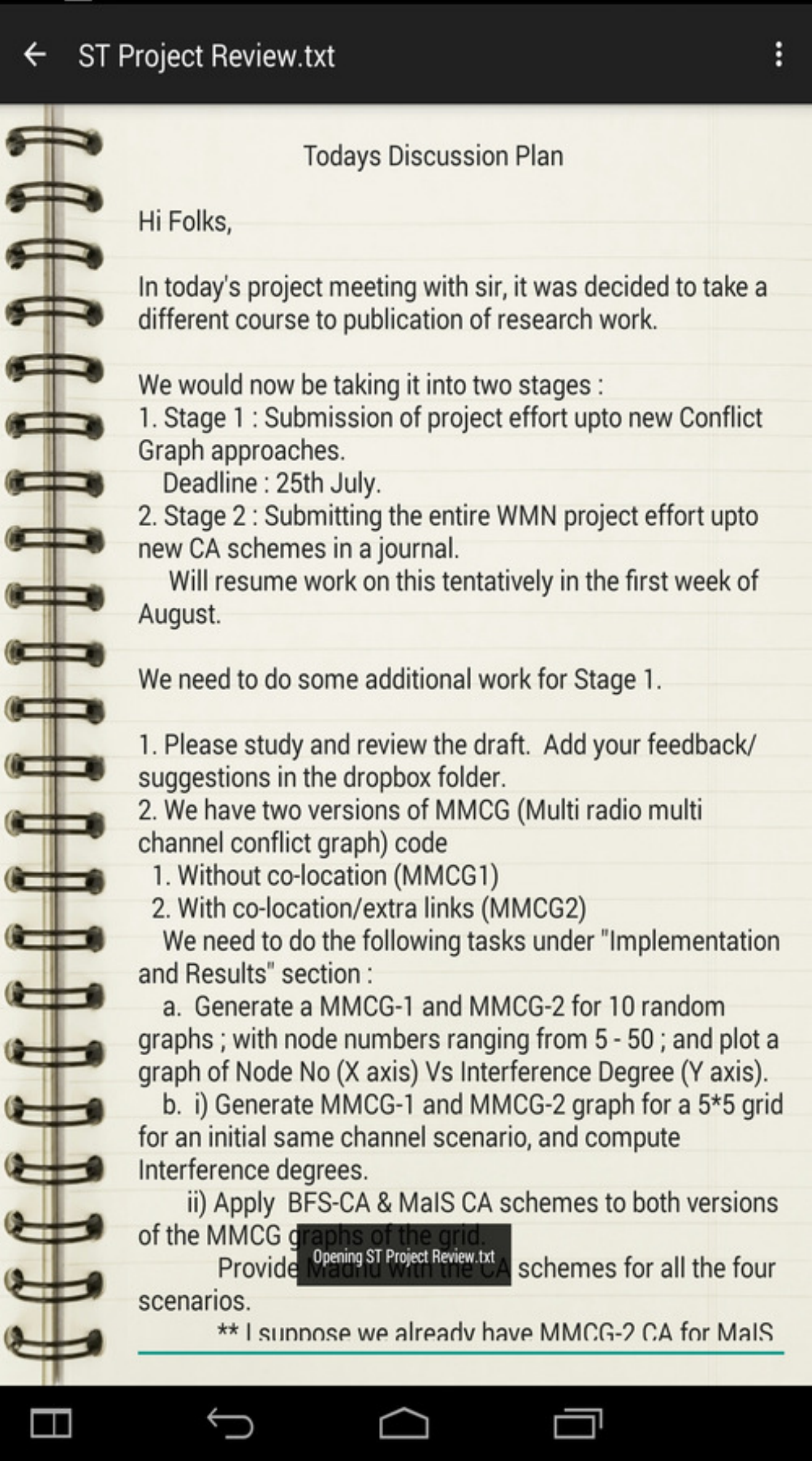}}\quad%
   \subfloat[Viewing 'Shared MoMs' File] {\includegraphics[width=4.25cm, height=6.5cm]{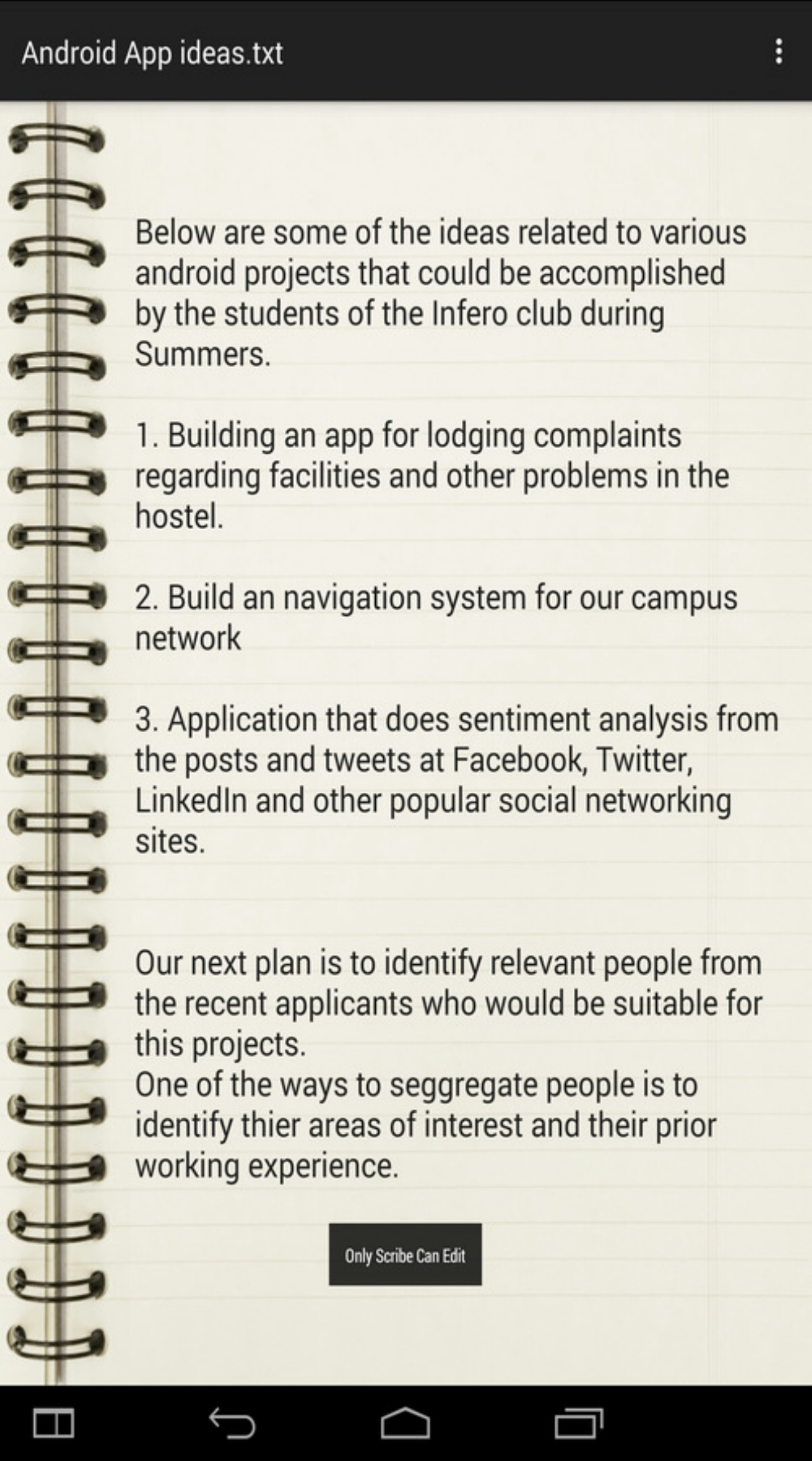}}
   \end{tabular}
    \caption{Editor Functions In Min-O-Mee} 
     \label{ss4}
\end{figure}

The text editor is a vital component of Min-O-Mee as the Scribe needs to take the MoM during the course of a meeting. The editor is available for editing only to the Scribe and not to any other Member. However, Members can either view the MoM being shared in real-time or the MoM file shared by the Scribe after compilation. In either case, a Member is not allowed to edit the MoM file. The text-editors for the two lists are depicted in Figure \ref{ss4}. It is evident from the Figure \ref{ss2} that the file 'ST Project Review.txt' is a part of the 'My MoMs' list of Lenovo Tablet, while the file 'Android App Ideas.txt' belongs to the 'Shared MoMs' list of the Samsung Galaxy smartphone. When a user tries to edit the file 'Android App Ideas.txt', it displays a \textit{toast} on the screen with the message 'Only Scribe Can Edit', as can be seen in Figure \ref{ss4} (b). In addition, the editor performs a periodic auto-save in the background and when the user hits the back button it again saves all the contents to 
the 'My MoMs' folder in the application installation directory. 

\begin{figure}[htb!]
                \centering
                \includegraphics[width=8.5cm, height=5cm]{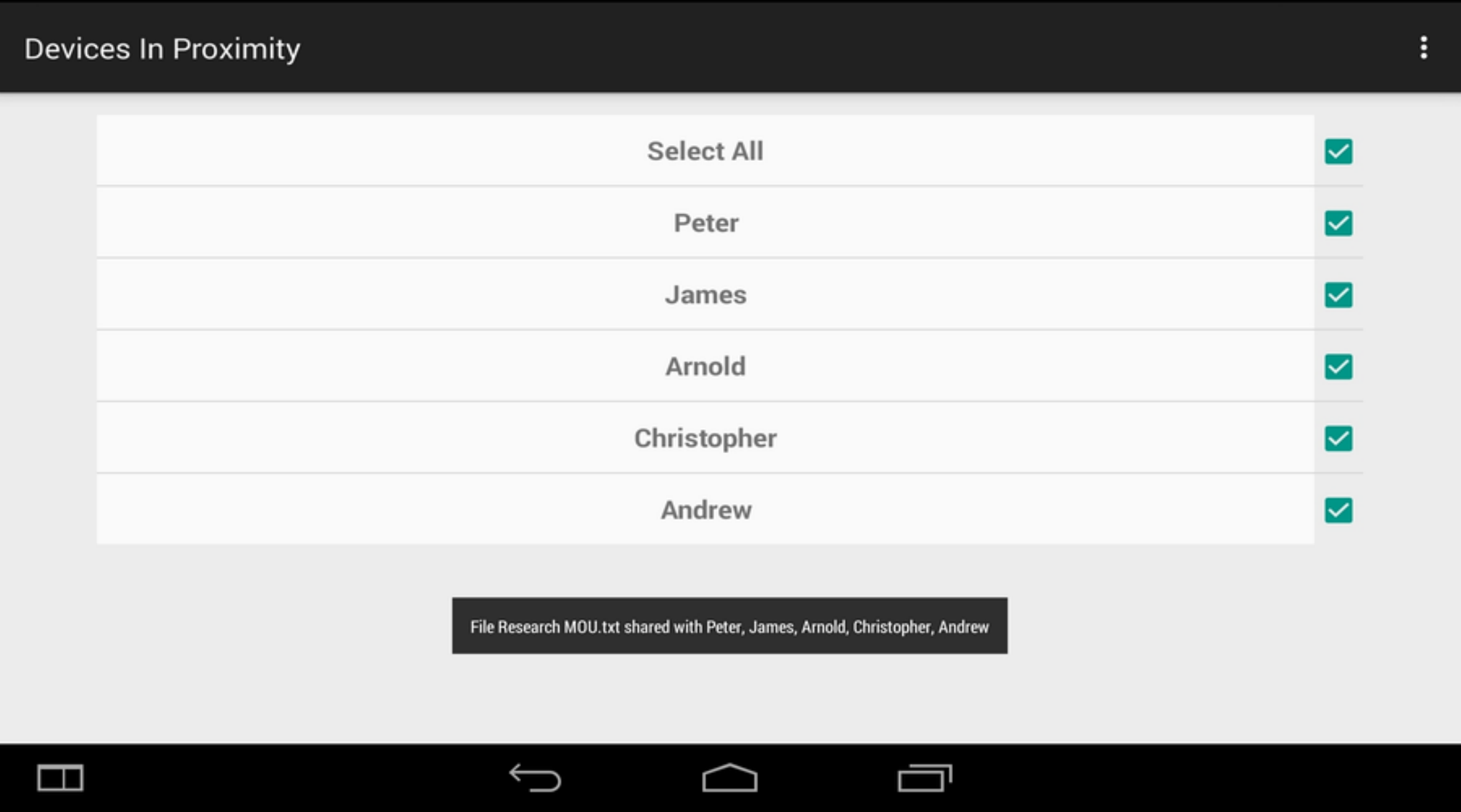}
                \caption{Scribe Sharing A MoM}
                \label{ss5}
        \end{figure}
\subsubsection{Data Sharing}
The objective of Min-O-Mee is to share the MoM with the Members in real-time. Deferred sharing entails that the Scribe should send a text/PDF file of the compiled MoM with all the Members at the end of the meeting. While in Real-Time sharing, content is updated in real time whenever it is modified by the Scribe. We facilitate Real-Time sharing in our implementation. 

Let us consider Figure \ref{ss5}, where the Scribe \textit{BRUCE} operating the Lenovo tablet, creates a MoM named 'Research MOU.txt'. To share the MoM with the Members, Min-O-Mee creates a list of devices in close proximity obtained through the background discovery mechanisms of the AllJoyn framework. The App offers BRUCE a choice to either share the MoM with a few selected Members or the entire Member list. A share button pops up which upon being pressed shares the MoM with the Members selected by BRUCE. The App generates a \textit{toast} notification stating the social names of the Members with whom the MoM has been shared.        

A Member which was selected by BRUCE to receive the MoM gets a notification message from the application about the incoming MoM file. The Member may choose to accept or reject the file. However, a Member which receives a MoM is not permitted by Min-O-Mee to share it with anyone else which can be inferred from the absence of \textit{share} option in Figure \ref{ss3} (b). Views of the 'My MoMs' and 'Shared MoMs' tab after sharing on 2 different devices can be seen in Figure~\ref{ss2}. Here the file 'Research MOU.txt' lies in the My MoMs' list of Lenovo Tablet (operated by Scribe:BRUCE) and in the 'Shared MoMs' list of the Samsung Galaxy DUOS I9082 smartphone (operated by a Member).

\begin{figure}
  \begin{tabular}{cc}
   \subfloat[Before Sharing]{\includegraphics[width=4.25cm, height=6.5cm]{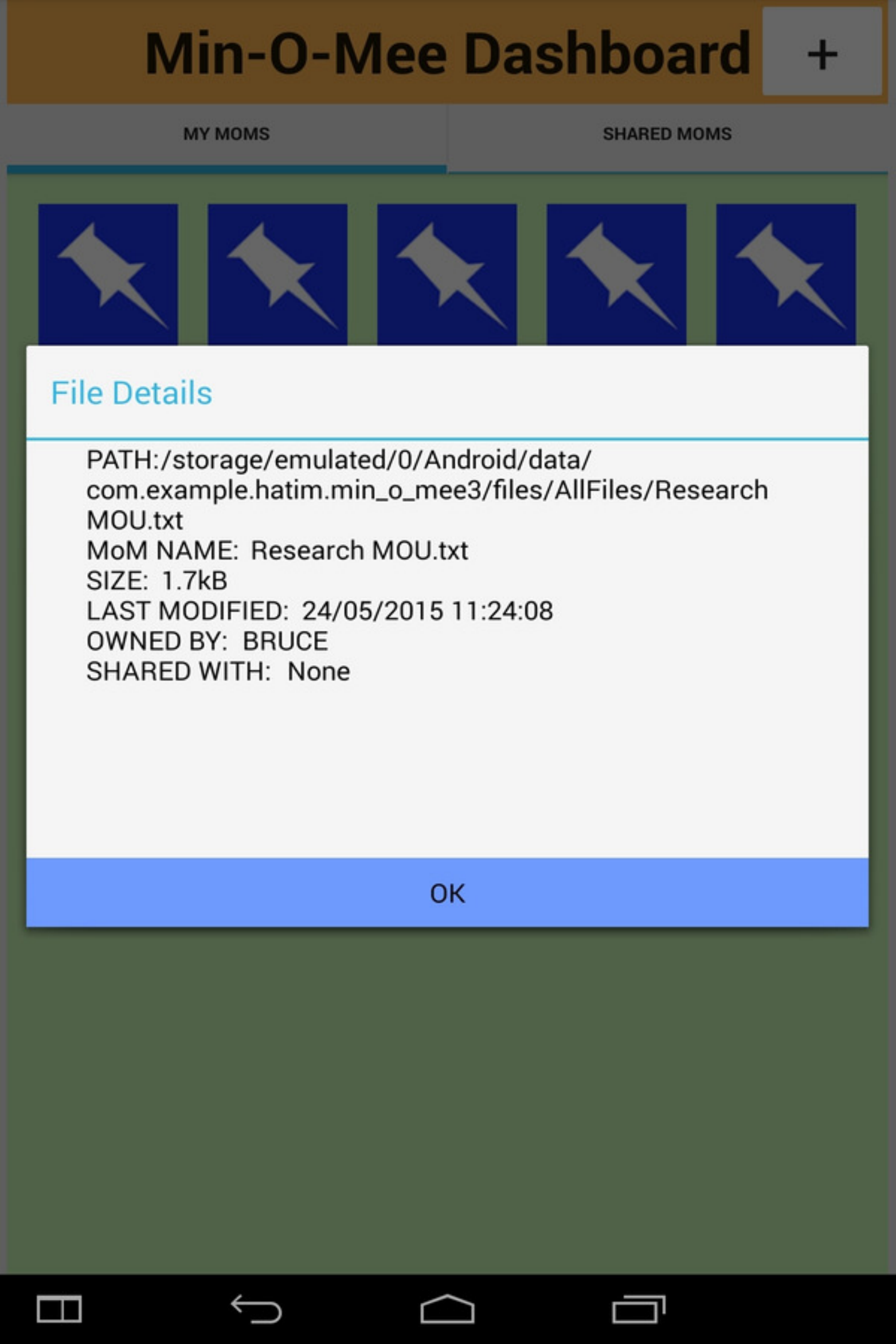}}\quad%
   \subfloat[After Sharing] {\includegraphics[width=4.25cm, height=6.5cm]{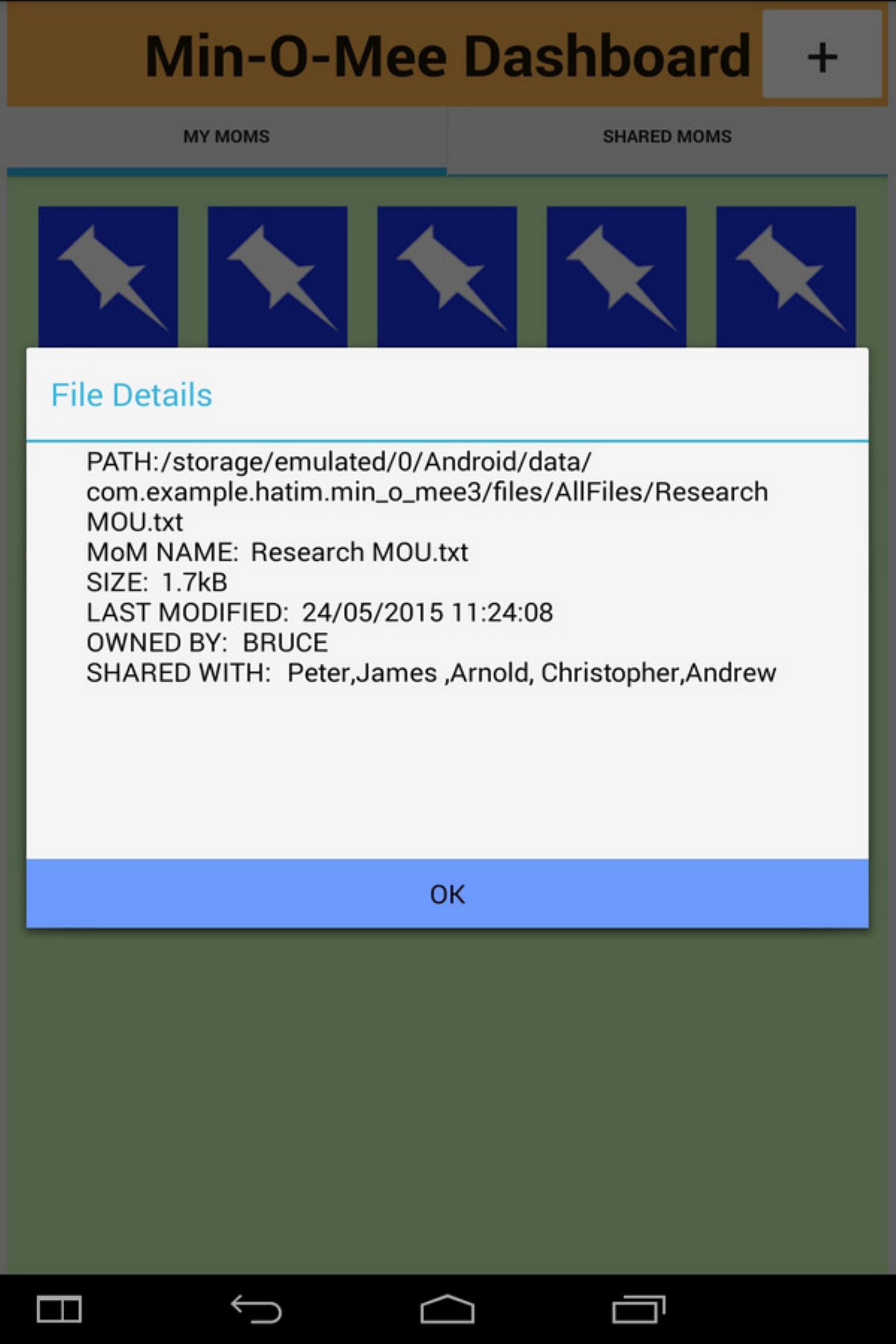}}
   \end{tabular}
    \caption{MoM File Attributes In Min-O-Mee} 
     \label{ss7}
\end{figure}

One important aspect of a shared MoM is the file metadata. Min-O-Mee keeps a record of file attributes, most important among which is  \textit{OWNED BY} which declares the owner of the MoM file. The owner is always the Scribe who created the MoM, regardless of its presence in the Scribe's device under the 'My MoMs' list, or in a Member device listed under 'Shared MoMs'. Thus the \textit{OWNED BY} attribute always remains constant. However, the attribute \textit{SHARED WITH} of a MoM file is modified by Min-O-Mee after the Scribe shares it with other Members. This can be noticed in Figure \ref{ss7}, which demonstrates that the \textit{SHARED WITH} attribute of a MoM file in the Scribe device is updated with the names of the Members after the file is shared with them. The underlying motivation of this exercise is to keep a track of the Members with whom the MoM is shared. 

\vspace{-1ex}
\section{\textbf{A Novel routing mechanism for PBNs}}
\vspace{-1ex}
We now illustrate the potential scenarios specific to the application where in multi-hop communication would be required and will be highly beneficial.
\begin{itemize}
 \item Consider an ongoing Min-O-Mee session where in there is a scribe and certain number of participants connected to the session hosted by the scribe. There might arise a scenario where in the scribe is mobile although with a limited speed. Due to his movement certain participants may fall outside his communication range. Thus the participants which fall outside the current range of the scribe may not receive the messages (updated content of an existing MoM or content of a new MoM) from the scribe. Communication range here refers to the Wi-Fi or Bluetooth range of a certain device.
\item Meeting or discussion between the members spans across multiple rooms on a floor or multiple floors. In this scenario every member of the meeting may not necessarily fall in the wireless communication range of the scenario.
\item A new participant joins the meeting at a some later point of time. But his current position does not fall in the range of the peer who is currently acting as a scribe. 
\end{itemize}

Multi-hop communication would serve as a prime solution for above scenarios. Now we describe about how we leverage the basic framework and extend it to support multi-hop communication 
\begin{figure}[htb!]
                \centering
                \includegraphics[width=7.1cm, height=5.2cm]{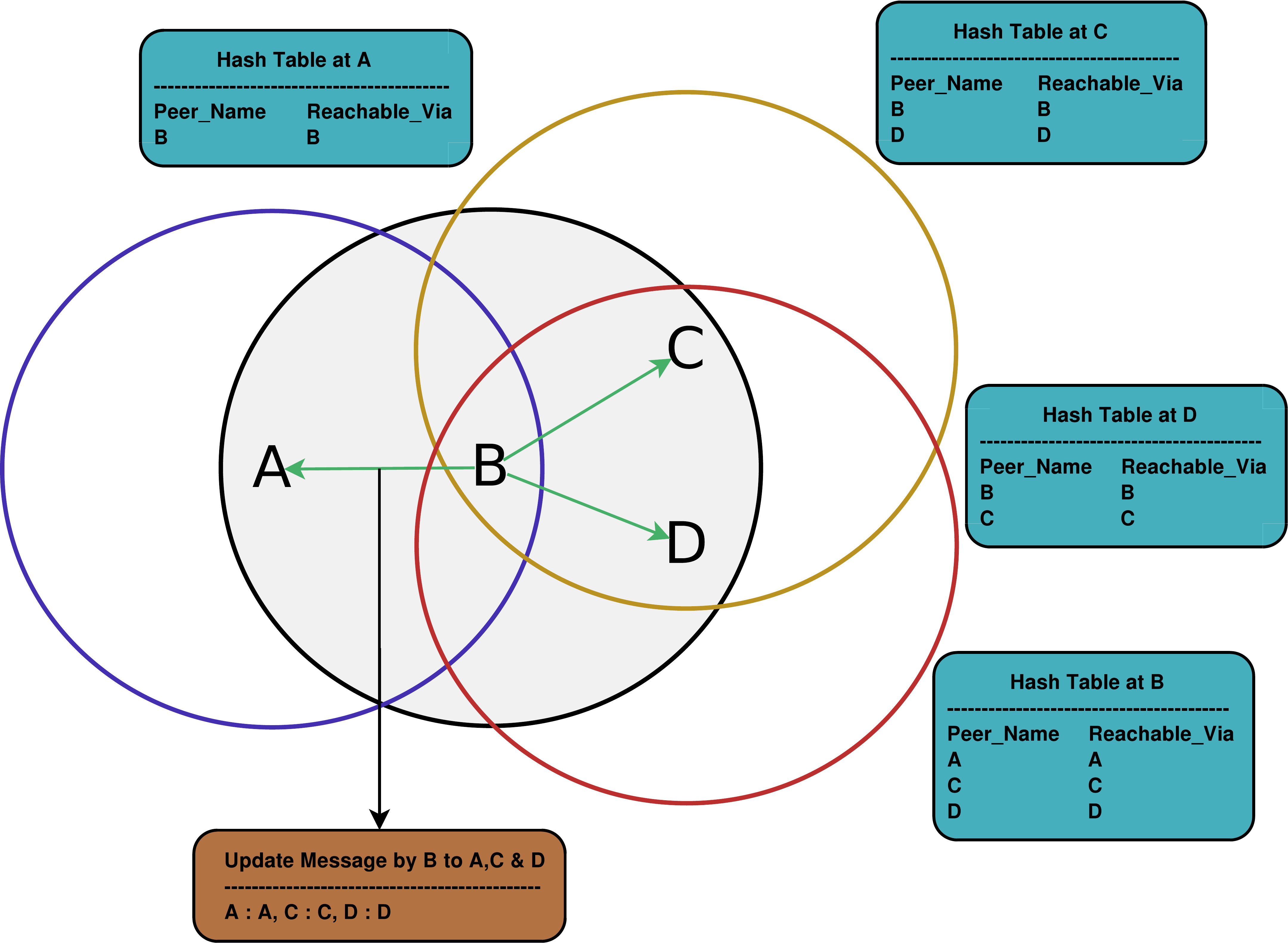}
                \caption{Routing in a sample PBN}
                \label{wrf}
        \end{figure}
\vspace{-1ex}
\subsection{Routing Functionality}
\vspace{-1ex}
In order to support multi-hop communication we take inputs from the traditional \textit{distance vector routing} (DVR) algorithm \cite{DVR} in the literature. In DVR each node (hop) periodically updates its nearby nodes with a distance vector. Distance vector from a node comprises of the distances to reach its nearby nodes. Through this a node gets to know about the other nodes which are reachable via its neighboring nodes. Based on the distance vector received from a neighboring node, each node determines which hop to choose as a next hop for a certain destination and updates its routing table accordingly.

In our implementation each node updates the nodes in the proximity with the list of nodes in its proximity. Consider for example scenario depicted in the Figure \ref{wrf}, where in node B has nodes A, C and D in its proximity. Through the information from node B, Node A gets to know that C and D are reachable via B and similarly the mechanism for nodes C and D. Details about the Hash Table and the Update Message from B in the figure is described in further discussion.

However, we do not include the distance vector in the update messages as in the DVR algorithm. Our prime goal is to facilitate a peer to discover its multi-hop peers through its neighboring peers. Also we assume that the members in the meeting are not highly mobile else otherwise the updates for topology change would be too frequent. This would take a large time for the routing information in the network to converge and frequent updates and processing of the messages would  drain the battery which is not ideal for battery powered mobile devices. Following points summarize the functional details of our routing algorithm implementation.
\begin{itemize}
 \item  In order to facilitate a peer communicate the list of its neighboring peers to the neighboring peers, we exploit the \textit{sessionless signal} available in the AllJoyn framework \cite{asa1}. Sessionless signals are signals that are broadcasted by a peer to all its reachable peers in the proximal network. This differ from session based signals which are sent only to participant(s) connected over a point-to-point or a multi-point session. Sessionless signals serves as a means for an application to publish some useful data in the network without prior establishment of a session.
 
 \item Initial Exchange of Routing Information when nodes arrive in each others proximity:
 Each node in our application advertises about its presence to others and at the same time discovers other nodes. This is achieved through advertisement and discovery mechanism in the AllJoyn router explained in the previous section. Each node maintains a list of discovered peers determined through discovery mechanism. We refer to this list as Routing Table or Hash Table in our further discussion. Whenever a peer obtains the information about a new discovered peer through AllJoyn router it updates its Routing Table and broadcasts an update message through sessionless signal to all its neighboring peer(s). Routing Table basically has following information :
\begin{itemize}
 \item Key : Peer\_Name.
 \item Value : Reachable\_Via.
\end{itemize}
Presence of a Peer\_Name (say X) in the Hash Table of a peer (say Y) indicates that peer Y is reachable from X. The corresponding value (Reachable\_Via, say Z) in the Hash Table concludes that Y is reachable from X via Z. If Y = Z then Y is the immediate neighbor of X.
Routing Table by a peer is maintained and updated as follows (Note: Symbols X, Y, Z or A, B, C denote Peer\_Names).
\begin{enumerate}

 \item  A Peer\_Name (say Y) is discovered by a peer named X and Y does not exist in the Hash Table of X. A new entry of the form (Y,Y) is inserted in the Hash Table of X. This basically indicates that Y is reachable and is the immediate neighbor of X.
 \item A Peer\_Name (say Y) is discovered by a peer named X and Y already exists in the Hash Table of X. In this case the value corresponding to the key Y in the Hash Table of X is only updated when the value is not equal to Y. If value corresponding to key Y is already Y then it conveys that Y is the immediate neighbor of X and there is no need to update the Reachable\_Via information for Y. If value is not equal to Y then the value for the peer Y is updated with the name of peer through which discovery message of Y was received. This scenario might arise when there is a peer (say A) which is reachable via multiple peer(s). In these circumstances we maintain the value of Reachable\_Via for A in the Hash Table with the name of peer through which discovery information about A was received at the last. We can conclude that an entry \textless A, B\textgreater in the Hash Table, where A = B represents a single hop path and it represents a multi-hop path if A != B.
\end{enumerate}
\begin{figure}[htb!]
                \centering
                \includegraphics[width=7.5cm, height=5cm]{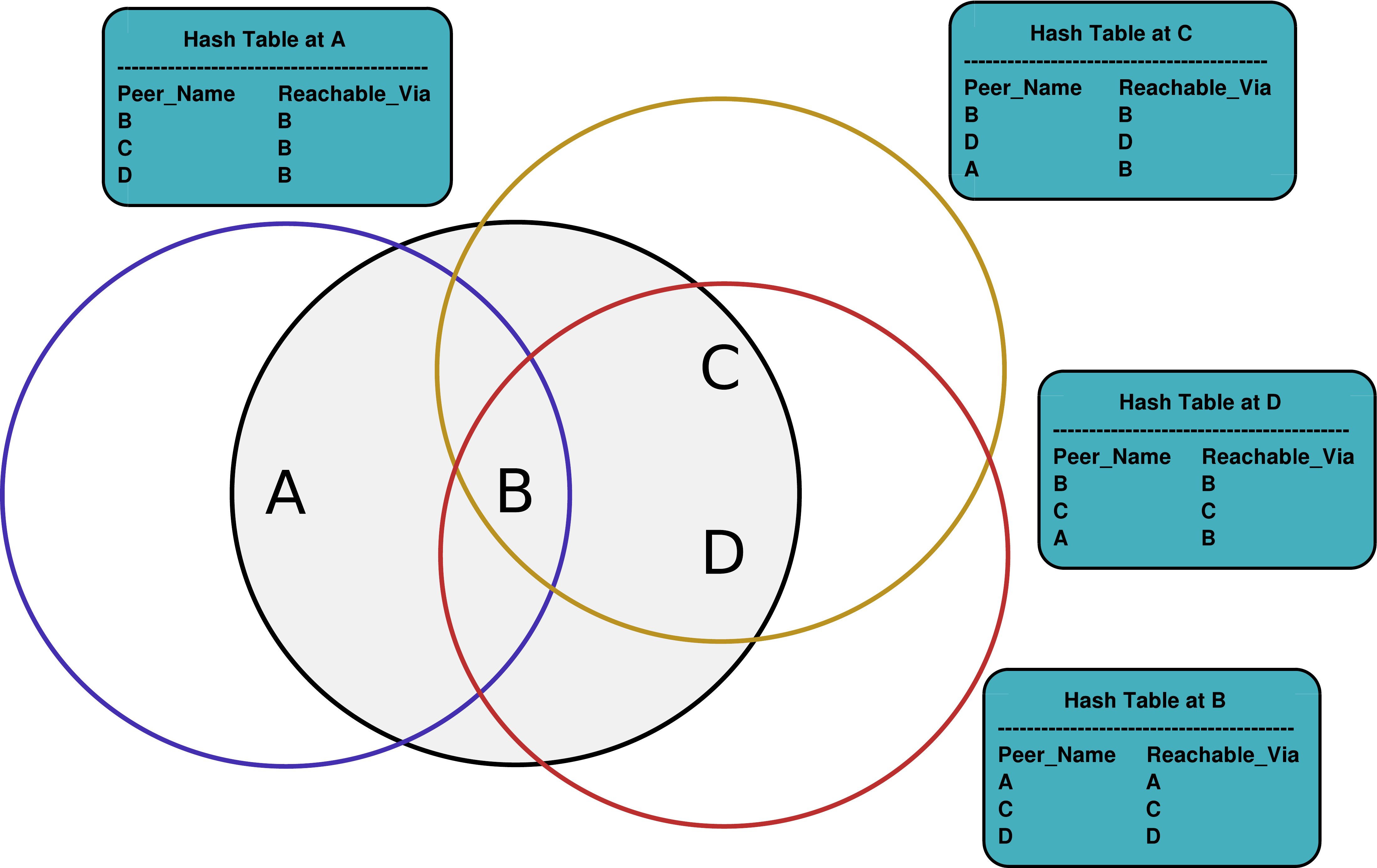}
                \caption{Hash Tables after processing of Routing update message}
                \label{update}
        \end{figure}

In context with Figure \ref{wrf}, consider at time t, nodes A, B, C and D in the figure knew about its immediate peers through discovery mechanism. Accordingly the Hash Tables at each of the nodes has the information as shown in the Figure \ref{wrf}. At this instance, B broadcasts an update message to all its immediate peers. This update message includes the current Routing Table at B. Hash Tables at each of the nodes after the processing of update message from B through above algorithm is shown in Figure \ref{update}. 

\item  Update on non reachability of an immediate peer:
AllJoyn router on each peers device periodically receives notifications regarding the presence of near by peers. It helps in determining the peers in the vicinity at any particular moment. A notification is generated by the router to the application on arrival of a new peer in the vicinity or on non availability of a peer which existed in immediate proximity at some earlier time. Thus when the application receives a notification about the non availability of a peer (say X), the routing table is updated accordingly and the entry corresponding to key X is deleted. This would trigger an update message comprising of the existing routing table to all the immediate neighbors of a peer. We do not maintain a timeout time in relation to the entries in the routing table due to the following reasons: 
\begin{itemize}
 \item AllJoyn router periodically receives information about the peers in the vicinity. It helps in determining the arrival of new peers or non availability of a previously existing peer in the vicinity. This would periodically update the routing table with new insertion/deletion or modifications and the table would not be stale.
 \item For multi-hop entries in the Hash Table (i.e the entries of the form \textless A, B\textgreater where A != B) of a peer, the peer would receive the non reachability information about A from B. This could be interpreted from the absence of Peer\_Name A in the routing update from B. Therefore we delete the entry \textless A, B\textgreater from the routing table, as A was not present in the routing update from B. Here, there could be a scenario where in A is reachable via other peers too, but the reachability information from B was received last. This resulted in the entry \textless A, B\textgreater in the Hash Table according to the algorithm mentioned previously. The absence of Peer\_Name in the list from B may indicate that A is not reachable via B. Peer named A may be reachable from other peers but since we had reachability information to A only via B and in the absence of A in the update message from B, we delete the entry \textless A, B\textgreater. In such cases the peer would receive reachability 
information of A from other peers if A is reachable through them and then a new entry could be inserted into the Hash Table. This primarily keeps the routing table updated.
\end{itemize}

\end{itemize}
\vspace{-1.2ex}
\section{Application Deployment And Testing}
\vspace{-1.2ex}
To test the performance of Min-O-Mee, the '.apk' file generated in the Android Studio IDE for the Min-O-Mee App was installed in five android devices. The minimum android version that is supported is 4.0 and the highest compatible version is 5.0. The five android devices we used are Lenovo Yoga Tablet, Samsung Galaxy DUOS I9082, Moto G, Micromax Unite 2 and HTC Desire 310. The App was successfully installed in all the devices and each device could discover the other four by making use of the AllJoyn discovery mechanism. A Wi-Fi hotspot is created on the Lenovo Tablet and all other devices joined the hotspot. We are able to create and share MoMs by assigning the role of Scribe to one of the devices and keeping the remaining devices as Members. The screenshots of the MoMs created and shared are presented in the previous section. We also tested the data delivery through multiple hops to the peers not in the immediate proximity of the Scribe.

\vspace{-1ex}
\section{Conclusion and Future Work}
\vspace{-.5ex}
In this paper we developed a MSNP application: Min-O-Mee for real time exchange of minutes-of-meeting between peers involved in a certain discussion/meeting. We provided an extenstion to the basic framework for faciliating multi-hop data transmission in the application so developed. In future, we plan to integrate other routing algorithms such as Link State Routing algorithm and routing algorithms in MANETS for high mobility scenarios. This would help in performance evaluation of different algorithms in different mobility scenarios with respect to metrics such as energy consumption, memory usage, convergence time, message overhead, etc.
 \vspace{-0.8ex}
 \section*{ACKNOWLEDGMENT}\label{p4}
 \vspace{-1ex}
This work was supported by the Deity, Govt of India (Grant No. 13(6)/2010CC\&BT).

\vspace{-0.8ex}
\bibliography{ref}

\end{document}